\documentclass[preprint2]{aastex}

\usepackage{amssymb,bm,graphicx,natbib}

\shorttitle{lensing in TeVeS}
\shortauthors{Chiu et al.}

\begin{document}

\title{\textbf{Theoretical Aspects of Gravitational Lensing in T$e$V$e\,$S}}
\author{Mu-Chen Chiu}
 \affil{Department of Physics, National Central University,
           Chung-Li, Taiwan 320, R.O.C.}
 \email{chiumuchen@gmail.com}

\author{Chung-Ming Ko}
\affil{Institute of Astronomy, Department of Physics and
          Center for Complex System, National Central University,
          Chung-Li, Taiwan 320, R.O.C.}
\email{cmko@astro.ncu.edu.tw}

\author{Yong Tian}
\affil{Department of Physics, National Taiwan University,
            Taipei, Taiwan 106, R.O.C.}
\email{yonngtian@gmail.com}

\date{submitted: ; accepted:}


\begin{abstract}

    Since Bekenstein's (2004) creation of his Tensor-Vector-Scalar theory (T$e$V$e\,$S),
    the Modified Newtonian dynamics (MOND) paradigm
    has been redeemed from the embarrassment of lacking a relativistic version.
    One primary success of T$e$V$e\,$S is that it provides an enhancement of
    gravitational lensing, which could not be achieved by other MONDian theories.
    Following Bekenstein's work, we investigate the phenomena of gravitational
    lensing including deflection angles, lens equations and time delay.
    We find that the deflection angle would maintain its value while the distance
    of closest approach vary in the MOND regime. We also use
    the deflection angle law to derive magnification and investigate microlensing light curves.
    We find that the difference in the magnification of the two images in the point mass model
    is not a constant such as in GR. Besides, microlensing light
    curves could deviate significantly from GR in the deep MOND regime.
    Furthermore, the scalar field,
    which is introduced to enhance the deflection angle in T$e$V$e\,$S, contributes
    a negative effect on the potential time delay. Unfortunately this phenomenon is unmeasurable
    in lensing systems where we can only observe the time delay between two images
    for a given source.  However, this measurable time delay offers another
    constraint
    on the mass ratio of the dark matter and MOND scenarios, which in general differs from that
    given by the deflection angle. In other words, for a lensing system, if two masses,
    $m_{gN}$ and $m_{gM}$, are mutually alternatives for the deflection angles in their own paradigm,
    regarding the time delay they are in general in an exclusive relation.

\end{abstract}


    \maketitle

    \section{Introduction}

    While Newtonian gravitational theory was applied to the
    extragalactic region, it no longer passed the trials as easily
    as it had done in solar system. Within Coma and Virgo cluster the velocities
    of individual galaxies are so large that its total mass must
    exceed the sum of masses of its shining galaxies if the clusters
    are gravitational bound systems \citep{Zwicky,Smith}. The 21 cm emission line from HI
    of spiral galaxies also show the need of extra masses besides
    visible stars \citep{vanAlba85,Bgmn89}. Moreover, X-ray observations and strong
    gravitational lensing in early type galaxies and clusters imply
    that there are $60 \sim 85\%$ masses in some undetected
    form~\citep{Fabb89,Bohringer95,Loew99}.
    This so-called ``missing mass" problem may be
    explained in two different ways: the existence of Dark
    Matter, and  a modification of Newton law
    , hence of General Relativity (GR). The most famous representative
    of the modified gravity theory party is Modified Newtonian
    Dynamics (MOND) proposed by Milgrom in 1983~\citep{M1,M2,M3},
    which assumes that Newton's second law needs
    to be modified when the acceleration is very small:
    \begin{equation}
    \tilde\mu(|\mathbf{a}|/\mathfrak{a}_0)\mathbf{a} =
  -\mbox{\boldmath{$\nabla$}}\Phi_{\rm N}.
    \label{MOND}
    \end{equation}

    Here $\Phi_{\rm N}$ is the usual Newtonian potential of the
    visible matter and $\mathfrak{a}_0\approx 1\times 10^{-10}\;$
    m s$^{-2}$ from the empirical data, such as the Tully-Fisher relation and rotation curves
    \citep{sv98};
    $\tilde\mu(x)\approx x$ for $x\ll 1$, $\tilde\mu(x)\rightarrow 1$ for $x\gg 1$.
    In the solar system where accelerations are strong
    compared to $\mathfrak{a}_0$, the formula (\ref{MOND}) is just
    Newton's second law $\mathbf{a} =-\mbox{\boldmath{$\nabla$}}\Phi_{\rm N}$.

    From this simple formulation the Tully-Fisher relation and flat rotation
    curves follow trivially~\citep{SandersMcGaugh}. The rotation curves of spiral galaxies, especially of LSBs,
    can be explained extremely well~\citep{BBS,rhs96,mcdb98,sv98}.
    Moreover, the dearth of dark matter in some
    ordinate elliptical galaxies can be intrinsically explained in MOND paradigm~\citep{MS03}.
    Even though many such phenomena in galaxy systems can be explained
    by this simple modification with an astonishing precision,
    the MOND paradigm had lacked a flawless relativistic gravitational
    theory for two decades~\citep{BekMilg,phase,SPCG,rhs97}, and thus could not
    address the
    lensing and other relativistic gravitational observations. However,
    since the end of March 2004, the embarrassment has been removed
    by ~\citet{TeVeS}, when he finally contrived a Lorentz covariant
    relativistic gravitational theory called T$e$V$e\,$S, which is based on
    three dynamical gravitational fields: an Einstein metric
    $g_{\mu\nu}$, a vector field $\mathfrak{U}_\mu$ and a scalar
    field $\phi$ (the acronym T$e$V$e\,$S is for this Tensor-Vector-Scalar
    content). In this scheme, T$e$V$e\,$S contains MOND for nonrelativistic
    dynamics, agrees with the solar system measurements for the
    $\beta$ and $\gamma$ PPN coefficients and avoids superluminal
    propagation of the metric, scalar and vector waves. Moreover, while concerning
    cosmogony, it not only has cosmological models which are similar
    to that of GR but also overcomes the Silk damping problem of the CDM-dearth universe with
    the help of the scalar field~\citep{SMFB05}. Now,
    we can investigate the relativistic MOND paradigm
    on many observations and
    experiments~\citep{gian05,SMFB05,zhao05a,zhao05b,zhaoeal05}.

    On the other hand, the GR with Dark Matter (DM) paradigm,
    despite its success for the large scale structure and the
    cosmic microwave background (CMB)(see e.g.~\citealt{spe03}), has some problems on the
    galactic scales. For example, the dark matter halos, if
    they really exist, must have a shallower core than the universal
    NFW density profile had predicted (cuspy core problem, see e.g.~\citealt{NS00,Rosky03});
     the disc size will be too
    small while considering the angular transformation from baryon
    to dark matter halo (angular momentum catastrophe, see e.g.~\citealt{NFW95, moore99, Maller02,kaza04}); the number of sub-halos predicted by
    non-baryon simulation is much more than the observed number of
    dwarf galaxies (sub-structure problems, see e.g.~\citealt{kauff93,NFW95}). Despite these problems,
    the GR with DM paradigm is less falsifiable and supported by a larger group
    because of its unlimited choice of models and parameters.

In order to judge which side better accounts for the phenomena of
the natural world, the disparity between T$e$V$e\,$S and GR with DM paradigms has to be closely
investigated. This distinction will be important
    for the physics of MOND and the ``missing mass" problem.
    Gravitational lensing (GL) phenomena provide us an opportunity
    because they dominate on scales that even surpass the
    size of dark matter halo~\citep{Koch04}.

    For a long time, light bending has been a great challenge for Modified Gravity.
    Any scalar field jointly coupling
    with the Einstein metric via a conformal transformation would yield the
    same optical phenomena as in the standard Einstein metric. Thus in the
    context of MOND, where little dark matter exists,
    it is hard to produce the anomalously strong deflection of
    photons~\citep{BekSan,Soussa1,Soussa2}. However,
    this problem can be solved by postulating the disformal relation between
    the physical metric and the Einstein metric~\citep{rhs97}.
    This is what Sanders' stratified theory, and then T$e$V$e\,$S take over.

    In this paper, we will discuss the theoretical aspects of
    Gravitational lensing systems within the framework of
    T$e$V$e\,$S. In \S~\ref{cosmology}, we summarize the characteristics
    of T$e$V$e\,$S and introduce a cosmology
    model within which gravitational lensing phenomena will be
    discussed. In \S~\ref{deflectionAngle} we derive the deflection
    angle in the general form and its two particular limits: the Newtonian and
    the MOND regimes. From this we obtain lens equations, which are the foundation of
    the other GL phenomena. In \S~\ref{timedelay}, we apply the
    standard time delay formalism in T$e$V$e\,$S. Finally,
    \S~\ref{discussion} summarizes the presuppositions of this
    paper and discusses some possible conclusions of our results in \S
    ~\ref{deflectionAngle} and \S~\ref{timedelay}.

    \section{Fundamentals of T$e$V$e\,$S and The Cosmological Model}\label{cosmology}
          Before introducing the action of T$e$V$e\,$S, we should
        explain some important features of the disformal metric relation.
        In Bekenstein's Tensor-Vector-Scalar theory (T$e$V$e\,$S), the
        role of Einstein's metric $g_{\alpha\beta}$ is replaced by the
        physical metric (Bekenstein 2004)
        \begin{equation}
        \tilde g_{\alpha\beta} = e^{-2\phi}(g_{\alpha\beta}+\mathfrak{U}_\alpha \mathfrak{U}_\beta) -
        e^{2\phi}\mathfrak{U}_\alpha \mathfrak{U}_\beta,
        \label{metricgeneral}
        \end{equation} with a normalization condition on the vector field:
        \begin{equation}
        g^{\alpha\beta} \mathfrak{U}_\alpha \mathfrak{U}_\beta = -1.
        \end{equation}  The physical metric, which governs the dynamical
        behavior of matter (the equivalence principle i.e. minimal coupling )
        is simultaneously influenced by
        a 4-vector field $\mathfrak{U}_\alpha$, a scalar field $\phi$
        and the Einstein's gravity $g_{\alpha\beta}$.

        From the theoretical viewpoint, if DM
        does not exist in galaxies or cluster scale, the
        concept of geodesics must be different from that of GR. Although
        there is no deeper theoretical base for the
        disformal relation~(\ref{metricgeneral}), empirical testament like lensing effects will be
        important to justify this choice~\citep{rhs97}.

        The cosmological model, which determines the background spacetime
        structure, is an essential ingredient in gravitational
        lensing~\citep{schn-ehl-fal92,PLW01}.
        On the one hand, the distance in the gravitational lensing
        scenario would vary under different cosmological models. On the
        other hand, gravitational lensing provides a tool to determinate
        $H_{0}$ and to constrain the cosmological parameters~\citep{BN86,schn96}.
        Since we have used an alternative gravitation theory,
        it is important to discuss its effects on cosmology,
        which is solely influenced by gravity. In this section, we introduce Bekenstein's
        Relativistic MOND theory, and then describe the revised Friedman model.

        \subsection{Actions of T$e$V$e\,$S}
            The convention in this paper are the metric signature $+2$ and units with
            $c=1$.

            There are three dynamical fields $g_{\alpha\beta}$,
            $\mathfrak{U}_\alpha$ and $\phi$ built in T$e$V$e\,$S.
            In addition to these fields, there is another non-dynamical scalar
            field $\sigma $, which is introduced into the scalar
            field action to eliminate the superluminal problem of
            $\phi$~\citep{TeVeS}.

            By varying the actions for these fields,
            the respective equations of motion for the scalar fields, vector
            fields and the Einstein metric can
            be arrived at, therefore the physical metric is determined.

            The action in T$e$V$e\,$S
            can be divided into four parts, $S=S_g+S_s+S_v+S_m$.
            The geometrical action,
            \begin{equation} S_g=(16\pi G)^{-1}\int g^{\alpha\beta}
            R_{\alpha\beta} (-g)^{1/2} d^4 x,
            \label{EH}
            \end{equation} is identical to the Einstein-Hilbert action in GR and
            is used to produce the Einstein metric, $g_{\alpha\beta}$. The matter action coupling
            to the scalar field, $\phi$, and vector field, $\mathfrak{U}_\alpha$, through $\tilde g_{\alpha\beta}$
            is taken to be
            \begin{equation} S_m=\int \mathcal{L}(-\tilde g)^{1/2}
            d^4x,
            \label{matter}
            \end{equation} where $\mathcal{L}$ is a lagrangian density for the fields under
            consideration and should be considered as
            functionals of the physical metric and its derivatives.

            The vector action with the form
            \begin{eqnarray} S_v &=&-{K\over 32\pi G}\int
            \big[g^{\alpha\beta}g^{\mu\nu}
            \mathfrak{U}_{[\alpha,\mu]} \mathfrak{U}_{[\beta,\nu]} \nonumber \\
            &\,&-2(\lambda/K)(g^{\mu\nu}\mathfrak{U}_\mu
            \mathfrak{U}_\nu +1)\big](-g)^{1/2} d^4 x
            \label{vector}
            \end{eqnarray} yields the vector field equation.
            Here, $K$ is a dimensionless parameter, and $\lambda$ is a Lagrange multiplier.
            Moreover, $\mathfrak{U}_{[\alpha,\mu]}$ means $\mathfrak{U}_{\alpha,\mu}
            -\mathfrak{U}_{\mu,\alpha}$.

            The action of the two scalar fields is taken to have the form
            \begin{eqnarray} S_s &=&-{\scriptstyle 1\over\scriptstyle
            2}\int\big[\sigma^2
            h^{\alpha\beta}\phi_{,\alpha}\phi_{,\beta} \nonumber \\
            &\,&+{\scriptstyle
            1\over\scriptstyle 2}G
            \ell^{-2}\sigma^4 F(kG\sigma^2) \big](-g)^{1/2} d^4 x,
            \label{scalar}
            \end{eqnarray}
            where $h^{\alpha\beta}\equiv g^{\alpha\beta}
            -\mathfrak{U}^\alpha \mathfrak{U}^\beta$;
            $F$ is a free function in order to produce
            the dynamical behaviors of MOND, which needs to be constrained
            empirically from cosmological model; $l$ is a constant length and $k$ is another dimensionless
            parameter in this theory. In the FRW-like cosmological models, $k$ is smaller
            than $10^{-2}$~\citep{TeVeS,SMFB05}.
 Variation of $\sigma$ of the action gives
\begin{equation}
 -\mu F(\mu ) -{\textstyle{1\over 2}}\,
 \mu ^2F^\prime(\mu ) = y \,
\label{F}
\end{equation}
where $y=k\ell^2 h^{\mu\nu}\phi_{,\mu}\phi_{,\nu}$ is a free
function of $\mu$ equivalent to $F$. However, in later discussions,
it is more convenient to consider $\mu(y)$ as a function of $y$.
Variation of $\phi$ of the action then gives
            \begin{eqnarray}
            \left[\mu\left(y\right)
            h^{\alpha\beta}\phi_{,\alpha} \right]_{;\beta}=  \; \;\;\;\;\;\;\;\;\;\; \;\; \;\;\;\;\;\;\;\;\;\; \;\; \;\;\;\;\;\;\nonumber \\
             \;\;\;\;\;\;\;\;
            kG\big[g^{\alpha\beta} + (1+e^{-4\phi}) \mathfrak{U}^\alpha
            \mathfrak{U}^\beta\big] \tilde T_{\alpha\beta}.
            \label{s_equation}
            \end{eqnarray}  As $\sigma$ through $\mu=kG\sigma^2$ can be expressed in terms of $\phi$ explicitly,
Eq.~(\ref{s_equation}) involves $\phi$ only.

             To illustrate ideas, in this work we adopt the free
function suggested by Bekenstein (2004)
\begin{equation}
 y={3\over 4}{ \mu^2(\mu-2)^2\over 1-\mu},
\label{y}
\end{equation}
whence $y$ approaches zero asymptotically as $y\sim 3\mu^2$ when
$\mu\rightarrow 0$, and $y\rightarrow\infty$ when $\mu\rightarrow
1$.

We are going to discuss gravitational lensing of a point mass in
\S~\ref{deflectionAngle} and \S~\ref{timedelay}. Thus we are
interested in static and spherically symmetric geometry. In this
case $\mu$ runs from 1 (the Newtonian regime) to 0 (the deep MOND
regime). In order to satisfy the behavior of the potential at these
limits we require $y\propto \mu^2$ asymptotically as $\mu\rightarrow
0$, and $y\rightarrow\infty$ as $\mu\rightarrow 1$. In the
intermediate regime $y(\mu)$ (or $\mu(y)$) is arbitrary.
Nonetheless, the exact functional form does not affect our results
because we do not consider the intermediate regime.

        \subsection{The Cosmological Model}
            As pointed out by~\citet{TeVeS}, the physical metric of FRW
            cosmology in T$e$V$e\,$S is obtained by replacing $dt \rightarrow e^{\phi}dt$
            and $a \rightarrow e^{-\phi}a$ in the primitive Roberson-Walker
            metric. Here, the Friedmann equation becomes:

            \begin{equation}
            \ {\dot{\tilde{a}}\over \tilde{a}}= e^{-\phi} \left( {\dot{a} \over
            a}-\dot{\phi}\right),
            \label{Friedmanneq}
            \end{equation} where $\dot{\tilde{a}}/\tilde{a}$ and
            $\dot{a}/a$ are essentially equal in the whole history because
            $e^{-\phi} \sim 1$ and $\dot{\phi} \sim 0$~\citep{TeVeS,hao05}.

            Since we have the additional scalar fields and vector fields
            in  Bekenstein's picture, the Einstein equation, which tells us the
            geometrical properties, is not only determined by
            the energy-momentum tensor as before, but also by the scalar and vector fields.
            Here, $\phi$, $\sigma$, and $\mathfrak{U}_{\alpha}$ are assumed to partake of the symmetry
            of the spacetime. The
            vector field is considered to be parallel with
            the cosmological time, i.e.
            $\mathfrak{U}^{\alpha}=\delta_t^\alpha$, because there should be no
            preferred spatial direction (Bekenstein 2004).
            Moreover, if we apply Weyl's postulate,
            which states that galaxies would move along nonintersecting world lines and
            therefore the spatial components in the energy-momentum tensor
            can be neglected~\citep{weyl23}, then only the $tt$ component of Einstein's equation
            remains:
            \begin{eqnarray}
            \lefteqn{{G_{tt}}={8\pi G}(\tilde\rho e^{-2\phi}+\sigma^2 \dot\phi^2)} \nonumber \\
            &&+ {2\pi\over k^2\ell^2}{\mu^2 F(\mu)+ \Lambda}
            \nonumber \\
            \lefteqn{={8\pi G}\tilde\rho e^{-2\phi}+{4\pi\over
            k^2\ell^2}} \nonumber \\
            & & \left[{\scriptstyle 3\over \scriptstyle
            2}\mu^2 F(\mu) + {\scriptstyle 1\over \scriptstyle
            2}\mu^3 F'(\mu) \right]+\Lambda. \label{Gtt}
            \end{eqnarray} The appearance of $\Lambda$ in Eq.~(\ref{Gtt}) results from
            the choice of the free function $F(\mu)$, which is decided by Eq.~(\ref{F})
            and Eq.~(\ref{y}). It is also possible to
            construct another form of $F(\mu)$ that makes the dark energy as
            an effective phenomenon of the scalar field $\phi$~\citep{hao05}. However, before having
            any theoretical reason for the choice of the free function $F(\mu)$, we would like
            to leave the interpretation alone and only treat it as a parameter
            in the cosmological model.




            By the Roberson-Walker metric, which determines the $tt$ component
            of the Einstein tensor Eq.~(\ref{Gtt}), we can get the modified
            Friedmann equation in a flat universe in the form
            \begin{eqnarray}  {\dot a^2\over a^{2}}={8\pi G\over 3}\rho_M
            +{8\pi G\over 3}\rho_{\phi}+{\Lambda \over 3}, \label{Friedmann2}
            \end{eqnarray} where ${\rho}_M=\tilde{\rho}e^{-2\phi}$ corresponds to
            the matter density in Einstein's picture and
            \begin{equation}
            \rho_{\phi}={3\over 2 G k^2\ell^2}\big[{\scriptstyle 3\over \scriptstyle
            2}\mu^2 F(\mu) + {\scriptstyle 1\over \scriptstyle
            2}\mu^3 F'(\mu) \big]
            \end{equation}
            is the density due to the scalar field. For convenience, we
            can rewrite Eq.~(\ref{Friedmann2}) in terms of $\Omega$ density:
            \begin{equation}
            \
            H^2=H_{0}^2[\Omega_{m}(z)+\Omega_{\phi}(z)+\Omega_{\Lambda}(z)].
            \label{Friedmann3}
            \end{equation} Since gravitational lensing must be observed
            within the framework of galaxies or clusters of galaxies,
            we only need to consider the cosmological model in
            the matter and $\Lambda$-dominated era, so $\Omega_{m}(z)
            \simeq \Omega_{m}(1+z)^{3}$. Furthermore, in order to avoid a significant effect
            of the integrated Sachs Wolfe term on CMB power spectrum, $\Omega_{\phi}(z)$
            must be extremely small compared with the matter
            density~\citep{SMFB05}. So even if we do not know exactly how
            $\Omega_{\phi}(z)$ evolves with $z$, we can neglect it, because its value
            is overwhelmed by the uncertainty of $\Omega_{m}$ and
            $\Omega_{\Lambda}$ for a non-CDM universe~\citep{mc99,mc04,SMFB05}.

            To summarize, in T$e$V$e\,$S the angular distance at any redshift
            can be calculated by
            \begin{equation}
            \\d_{A}\simeq{{(1+z)^{-1}} \over
            H_{0}}\int^{z}_{0}{dz\big[{\Omega_{m}(1+z)^{3}
            +\Omega_{\Lambda}} \big]^{-{1/2}}}.
            \label{angulardistance}
            \end{equation}
            We conclude that if there is any difference in the angular
            distance between GR and T$e$V$e\,$S,
            it arises from the low matter density universe in T$e$V$e\,$S rather than the
            influence of the scalar field $\phi$.

            Finally, it is worth noting that although it is not trivial
            to consider an inhomogeneous universe on the gravitational
            lens scale (Dyer $\&$ Roder 1973), we do not consider that case because we only deal with a
            single point mass lens embedded in a background Friedmann universe in
            this paper.

    \section{Gravitational Lensing in T$e$V$e\,$S}\label{deflectionAngle}
        When building gravitational lensing models, people usually use some approximations,
        such as the static and thin lens approximations. We also adopt the weak
        field assumption since all known gravitational lens phenomena are
        influenced by weak gravitational systems.
        Although the Schwarzschild lens (point mass model) is not sufficient
        for gravitational lensing effects,
        we consider this for its simplicity.
        Moreover it gives results of the same order-of-magnitude as those
        for more realistic lenses~\citep{schn-ehl-fal92}.

        \subsection{Deflection Angle in Terms of Mass}
            In the weak field limit, the physical metric of a static
            spherically symmetric system in T$e$V$e$S can be expressed in the isotropic form:
            \begin{eqnarray}  \lefteqn{\tilde{g}_{\alpha\beta}\, dx^\alpha
            dx^\beta= -(1+2\Phi)dt^2} \nonumber \\
            && +(1-2\Phi)[d\varrho^2+
            \varrho^2(d\theta^2+\sin^2\theta
            d\varphi^2)],  \label{weakspherical}
            \end{eqnarray} where $\Phi=\Xi\Phi_{N}+\phi$ and $\Xi\equiv e^{-2 \phi_{c}}(1+K/2)^{-1}$
            with $\phi_{c}$ as the asymptotic boundary value of $\phi$ and $K<10^{-3}$ from the constrain of
            PPN parameters (Bekenstein 2004).

            Considering a quasistatic system with a perfect fluid, the scalar
            equation ~(\ref{s_equation}) can be reduced into
            \begin{equation}
            \nabla\cdot\Big[\mu(y)
            \nabla\phi\Big]=kG\tilde\rho,
            \label{sca_eq}
            \end{equation} because $\mathfrak{U}_\alpha$ has only
            a time-independent temporal component and $\phi \ll 1$ in this
            case. Here $y=k\ell^2 (\nabla\phi)^2$, which is obtained from the general form,
            $y=k\ell^2 h^{\mu\nu}\phi_{,\mu}\phi_{,\nu}$, in
            spherical symmetric geometry.
            From Eq.~(\ref{sca_eq}), we can find a relation between $\Phi_{N}$ and $\phi$,
            \begin{equation}
            \nabla \phi=\left(k/4\pi \mu \right)\nabla \Phi_N.
            \label{dscalar}
            \end{equation} Therefore, $\Phi$ can be determined whenever $\mu$ is
            known.

            It is also worth to stress that
            $\tilde{g}_{tt}=-(1+2\Phi)$ follows the empirical definition of the
            gravitational mass $m_{g}$ in the solar system when $\mu\rightarrow 1$:
            \begin{equation}
            g={1 \over 2}{\partial \tilde{g}_{tt} \over \partial \varrho} =(\Xi
            +k/4\pi){G m_{g} \over \varrho^2}={G_N m_{g} \over \varrho^2},
            \label{metricsymmetric}
            \end{equation} and that the spatial components,
            $\tilde{g}_{ii}=(1-2\Phi)$, guarantee the theory passes the
            classical post-Newtonian (PPN) tests of gravity (Bekenstein 2004).

            We now apply the static spherically symmetric metric, c.f. Eq.~(\ref{weakspherical}),
            to the GL systems. Considering a light ray, which propagates
            on a null geodesic, and moves in the equatorial plane, i.e.
            $\theta=\pi/2$, we can arrive at three constants of motion
            from the physical metric in equation~(\ref{weakspherical})
            with some proper affine parameter, $\tau$:
            \begin{eqnarray}
            (1+2\Phi) \dot t=E, \label{1stcmotion}
            \\ (1-2\Phi) \varrho^2\dot\varphi=L, \label{2ndcmotion}\\
            \ (1-2\Phi)\dot{\varrho}^{2}+(1+2\Phi)\varrho^{-2}{L}^2
            -(1-2\Phi) E^2=0, \label{3rdcmotion}
            \end{eqnarray} where over dot denotes the derivative with respect to $\tau$.
            Eliminating E and L in Eq.~(\ref{3rdcmotion})
            from Eq.~(\ref{1stcmotion}) and ~(\ref{2ndcmotion}), and recalling the fact that
            $ \dot{\varrho}$ vanishes at the closest approach,
            $ \varrho=\varrho_{0}$, yields
            \begin{equation}
            \ b^2 \equiv {L^2 \over E^2}={\varrho_{0}^2(1-2\Phi_{0})\over
            (1+2\Phi_{0})}, \label{impact}
            \end{equation} where, $\Phi_{0}\equiv\Phi(\varrho_{0})$. Combining
            Eq.~(\ref{impact}) and the three constants of motion, gives the
            equation which describes the shape of the photon orbit:
            \begin{equation} -(1-4\Phi)+(1-4\Phi_0)(\varrho_0/\varrho)^2
            [\varrho^{-2}(d\varrho/d\varphi)^2+1]=0.
            \end{equation}
            The solution of this equation in quadrature is
            \begin{equation}
            \varphi=\int^\varrho{\left\{({\varrho \over \varrho_{0}})^2
            \left[1-4(\Phi-\Phi_{0})\right]-1\right\}^{-1/2}{d\varrho \over \varrho}}.
            \label{angle1}
            \end{equation} If we take the Taylor expansion of Eq.~(\ref{angle1}) to the first order of
            $\Phi$, then it is easy to see that the quadrature is a combination of the angle of
            a straight line feeling no gravity and the angle of the deviation due to the gravitation field.
            Hence the deflection angle can be approximated from the first order
            terms of this Taylor expansion:
            \begin{equation}
            \Delta\varphi=\varrho_{0} \int^{\varrho}_{\varrho_0}
            {{2\varrho(\Phi-\Phi_{0})} \over {\left(\varrho^2-\varrho_{0}^2\right)^{3/2}}}{d\varrho}.
            \label{angle2}
            \end{equation}
            Furthermore, after some manipulations (see e.g. Bekenstein 2004),
            the deflection angle is
            \begin{equation}
            \Delta\varphi=\varrho_{0} \int^{\varrho}_{\varrho_0}
            {4\Phi'\over (\varrho^2-\varrho_{0}^2)^{1/2}}\ d\varrho
            -\varrho_{0}{{4\Phi}\over{({\varrho^2-\varrho_{0}^2})^{1/2}}}.
            \label{angle3}
            \end{equation}

            It is obvious
            that the deflection angle between the closest approach and
            any position is always positive.
            First, gravity is always attractive (therefore $\Phi'>0$),
            so the value of the quadrature is positive because every term
            in it is positive. Second, the second term of the right hand side is
            negligible even in the MOND regime, in which $\Phi$ behaves as $\ln \varrho$,
            if $\varrho$ is large.

\begin{figure}[t]
\plotone{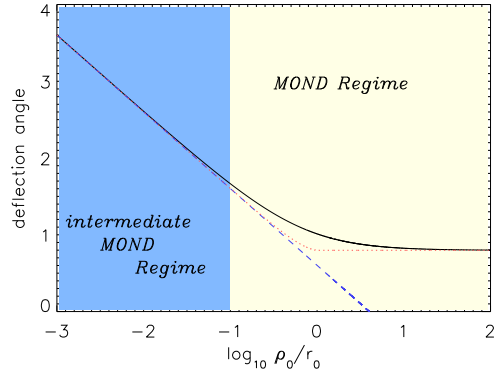} \caption{\label{figure1} Deflection angle of light
derived under the point mass model in T$e$V$e\,$S with $k=0.01$ and
$K=0.01$(solid line) and GR (dashed line). The closest approach is
normalized to the length scale $r_{_0}$, which depends on the mass
of the lens. The dotted line shows the work of Mortlock and Turner
(2001), who started from a reliable intuition without a relativistic
MOND. In the two extreme limits, their result is exactly the same as
the deflection law derived in T$e$V$e\,$S. However, in the
intermediate region there is a difference.}
\end{figure}

        \subsection{Two Limits of the Deflection Angle}
            Even though we can make sure that the deflection angle can be
            enhanced in T$e$V$e\,$S by the simple argument above, we
            need the exact form of the the potential to reach the real
            value of a deflection angle. Unfortunately, the form of potential
            depends on the free function $F(\mu)$, which can only be constrained empirically.
            However, there are two limits, the Newtonian and the MOND
            regime, in which the dynamical behavior has been
            clearly known, and so to the potential form. Actually, any
            choice of the $F(\mu)$ cannot contradict the empirical
            potential form in these two limits. Here, we discuss the
            deflection angles in the two regimes (see
            Fig.~\ref{figure1}).

            The MOND regime means $\nabla \phi$ is significant
            compared with $\nabla \Phi$, whence by Eq.~(\ref{dscalar}),
            $\mu \ll1$ has to be a sufficient condition for this regime.
           From Eq.~(\ref{y}) and $y=k\ell^2 (\nabla\phi)^2$,
            we can obtain $\mu \approx (k/3)^{1/2}l |\nabla
            \phi|$ when $\mu \ll1$ (i.e., in MOND regime). It has to be
            mentioned that although this relation results from
            the freedom of an arbitrary function, $\mu \sim
            |\nabla \phi|$ should always persist in order to guarantee $\nabla
            \phi$ varying as $1/\varrho$ in the deep MOND regime, a condition which
            follows from the MOND paradigm.
            Using this relation in Eq.~(\ref{dscalar}) to eliminate $\nabla \phi$ and
            replacing $\nabla \Phi_N$ with $\nabla \Phi$ gives
            \begin{equation}
            \ \mu=(k/8\pi\Xi)\big(-1+\sqrt{1+4|\Phi'|/\mathfrak{a}_0}\,\big)={k\over
            4\pi}{\left(|\Phi_N'|\over
            \Xi\mathfrak{a}_0\right)^{1/2}}.
            \label{mu}
            \end{equation} Here, the constant
            \begin{equation}
            \mathfrak{a}_0 \equiv{(3 k)^{1/2} \over 4\pi\Xi\ell}
            \end{equation}
            can be identified as Milgrom's constant. We should keep in mind that this form of
            $\mu$ is valid under the condition $\Phi' \ll (4\pi/k)^2\mathfrak{a}_0$.
            It also offers a criterion for distinguishing the Newtonian and the MOND regime.

            We define a distance,
            \begin{equation} \\r_0 \equiv \left(G_Nm_g\over\mathfrak{a}_0\right)^{1/2},
            \end{equation} at which acceleration, $\mathbf{a}$, equals to Milgrom's constant,
            $\mathfrak{a}_0$. Then a particle is in the
            deep MOND regime if
            $\varrho \gg kr_0/4\pi$, and in the Newtonian regime if $\varrho \ll kr_0/4\pi$.

            \subsubsection{The Newtonian regime}
                Although a photon moving in a GL system may
                come from a distance $\varrho \gg kr_0/4\pi $ to the closest
                approach $\varrho_0\ll kr_0/4\pi$ and then fly away again,
                the influence of gravity only dominates in the region
                which is very near the gravitation center.
                Therefore we call a GL system as being in the Newtonian
                regime if $\varrho_0\ll kr_0/4\pi$.

                In the Newtonian regime, $\Phi'$ in Eq~(\ref{angle3}) can be
                calculated from the relation $\Phi=\Xi\Phi_N+\phi$
                and Eq.~(\ref{dscalar}) for $\mu\rightarrow 1$. This would yield the
                deflection angle of a Schwarzschild lens,
                \begin{equation}
                \Delta\varphi={4Gm_g\over
                \varrho_0}\big(\Xi+{k\over 4\pi})={4G_Nm_g\over
                \varrho_0}.
                \label{angleN}
                \end{equation} Here again, $G_N$ is identical to the
                gravitational constant measured in a local
                experiment. This is the same as the result from GR.



            \subsubsection{The Deep MOND Regime}
                In the deep MOND regime, $\varrho_0 \gg kr_0/4\pi$, $\mu$ has the form of Eq.~(\ref{mu}).
                Then along with Eq.~(\ref{dscalar}) and Eq.~(\ref{mu}), the
                deflection angle for a point mass model in the deep MOND regime can be arrived
                at from Eq.~(\ref{angle3}):
                \begin{eqnarray}
                \lefteqn{\Delta\varphi={4G_N m_g\over
                \varrho_0(\Xi+k/4\pi)}} \nonumber \\
                & & \cdot\left\{\Xi+{\pi\over2}\varrho_0{\left[\mathfrak{a}_0(\Xi^2+k\Xi/4\pi)\over{G_Nm_g}\right]^{1/2}}\right\},
                \label{lenform}
                \end{eqnarray} with $\Xi\equiv1-K/2-2\phi_c$. Since all of
                $K$, $\phi_c$ and $k$ are much less than $1$, for a given mass, the
                deflection angle in the deep MOND regime differs from that
                in GR (or equivalence in the Newtonian regime; Eq.~[\ref{angleN}])
                by an amount almost independent of the distance of the
                closest approach.

                We have to stress that even though we do not know
                exactly how deflection angle varies with respect to
                the closest approach in the intermediate MOND
                regime, our result based on the deep MOND assumption
                (i.e., $\mu \ll 1$), which is incorrect in the regime
                $\varrho_0 \leq kr_0/4\pi$, approaches to the Newtonian
                prediction while  $\varrho_0$ decreasing (see
            Fig.~\ref{figure1}). Therefore
                we believe even in the intermediate regime,
                the deflection law based on the deep MOND
                assumption is approximately correct.

            \begin{figure}[t]
\epsscale{1.0} \plotone{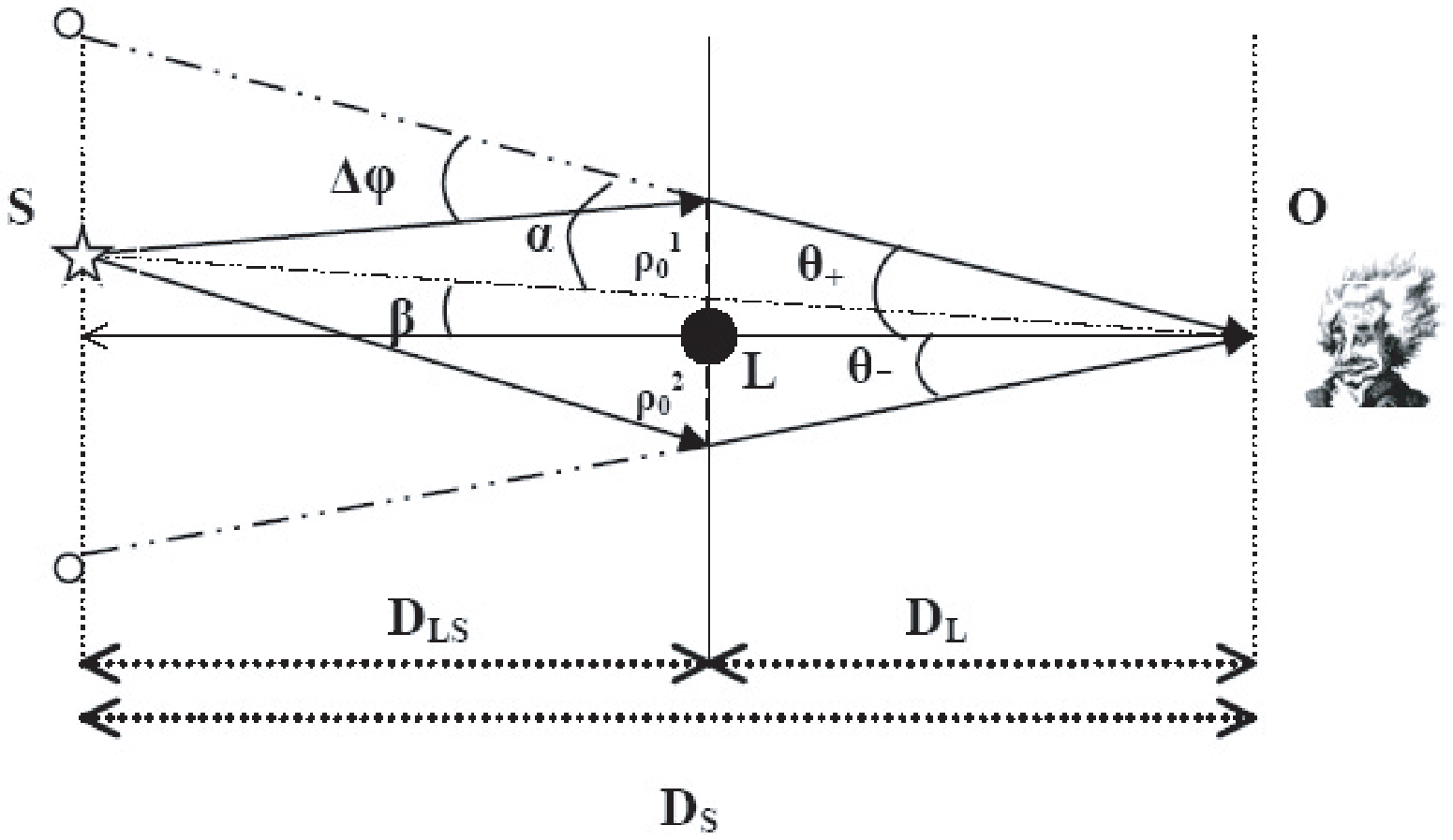} \caption{\label{figure0}
Embedding diagram of light bending. For a thin lens, the lensing
equation and geometrical time delay can be immediately read off. }
\end{figure}

        \subsection{Magnification and Microlensing}
            From the observational point of view, the measurable data is not deflection
            angles but positions of the projected sources, i.e. $\theta\equiv \varrho_0/D_L$
            . Here, $D_L$ is the angular distance of lens (Eq.~[\ref{angulardistance}]).
            Therefore we need a relation to connect
            $\theta$ and $\Delta\varphi$, the deflection angle derived in the last section.
            We can obtain their relation from an embedding diagram (Fig.~\ref{figure0}) and
            get the lens equations:
            \begin{eqnarray}
            \beta=\theta_{_+}-\alpha(\theta_{_+}),
            \label{lenequ}\\
            \beta=\alpha(\theta_{_-})-\theta_{_-}.\,
            \label{lenequ2}
            \end{eqnarray} Here the reduced deflection angle is defined as
            $\alpha(\theta)=(D_{LS}/D_S)\Delta\varphi(\theta)$. In GR it is not important
            to distinguish the difference between Eq.~(\ref{lenequ}) and Eq.~(\ref{lenequ2})
            in finding $\theta$~\citet{pac86}. It is, however,
            not the case in T$e$V$e\,$S. In order to keep
            $\theta_{_+}=\theta_{_-}$ when $\beta \rightarrow 0$, we
            must use Eq.~(\ref{lenequ}) for $\theta_{_+}$ and Eq.~(\ref{lenequ2})
            for $\theta_{_-}$.

            After substituting $\Delta\varphi$ and $\theta$ into the lens equation
            Eq.~(\ref{lenequ}), we get
            \begin{eqnarray}
            \beta=\theta- {D_{LS}\over D_LD_S}{4G_Nm_g\over
            \theta}     \nonumber \;\;\;\;\;\;\;\;\;\;\;\;\;\; \;\;\;\;\;\;\;\\
            \cdot\big[{\Xi \over \Xi+k/4\pi}+{\pi\over2}{\theta D_L\over
            r_0}(\Xi+{k \over 4\pi})^{-{1/2}}\big].
            \label{leneqnMOND}
            \end{eqnarray} For clarity, we can define
            $\theta_0\equiv r_0/D_L$ and $\theta_E$ with the form

            \begin{equation}
            \theta_E=\left(4G_Nm_g{D_{LS}\over {D_L D_S}}\right)^{1/2}.
            \label{thetaE}
            \end{equation} $\theta_E$ is called the Einstein
            radius; it is used as a length scale in the standard GL modeling.
            As $k\ll 1$ and $K\ll 1$, the lens equation ~(\ref{leneqnMOND})
            can be reduced to
            \begin{equation}
            \beta\simeq\theta-{{\theta_E}^2\over
            \theta}\big[1+{\pi\over2}{\theta\over
            \theta_0}\big]
            \label{graeq2}
            \end{equation} with the reduced deflection angle, $\alpha(\theta)$, as
            \begin{equation}
            \alpha\simeq{{\theta_E}^2\over
            \theta}\big[1+{\pi\over2}{\theta\over
            \theta_0}\big].
            \label{alphatheta}
            \end{equation} If Eq.~(\ref{lenequ2})
            is used instead of Eq.~(\ref{lenequ2}), then the only
            difference in Eq.~(\ref{graeq2}) is from $\beta$ to
            $-\beta$. It is easily seen that the effective
            contribution of the scalar field can be expressed as
            ${\pi}{{\theta_{_E}}^2 /2\theta_{_0}}$ and is
            independent of $\theta_0$ (or equivalently $\varrho_0$).

\begin{figure}[t]
\epsscale{1.0} \plotone{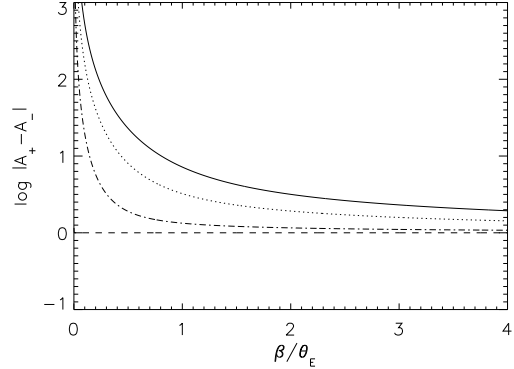} \caption{\label{mag} The
difference in the magnification of the two projected images. In
T$e$V$e\,$S, it deviates from the constant value in GR (dashed
line). From the solid to dotted-dashed line, they represent the
cases of T$e$V$e\,$S with $\theta_E /\theta_0 =$ 1.0, 0.5 and 0.1.}
\end{figure}

            For a point source, there are always two projected
            images, which can be obtained from Eq.~(\ref{lenequ})
            and Eq.~(\ref{lenequ2}). The solutions are
\begin{equation}
\theta_{\pm}=\pm{1 \over 2}\left\{\left(\beta\pm{\pi\theta_E^2 \over
2\theta_0}\right)\pm\left[\left(\beta\pm{\pi\theta_E^2 \over
2\theta_0}\right)^2+4\theta_E^2\right]^{1/2}\right\}>0.
\label{projimag}
\end{equation} Moreover, since gravitational lensing conserves surface brightness,
the magnifications of the image to source intensity are determined
by their area ratio,
\begin{equation}
A_{\pm}=\left|{\theta_{\pm} \over \beta}{\partial\theta_{\pm} \over
\partial\beta}\right|.
\end{equation} Interesting enough, the difference in the magnification of the two images
is no longer constant in T$e$V$e\,$S as in GR (see Fig.~\ref{mag}):
\begin{equation}
\left|A_{+}-A_{-}\right|\geq 1
\end{equation} And in the deep MOND regime, the total magnification is given by
\begin{eqnarray}
A\equiv A_{+}+A_{-}= \left(u_{1}\over
2u\right)\left[{{u_1}^2+2\over{u_1}\left
({u_1}^2+4\right)^{1/2}}+1\right] \\
+ \left(u_{2}\over 2u\right)\left[{{u_2}^2+2\over{u_2}\left
({u_2}^2+4\right)^{1/2}}-1\right] , \label{magtotal}
\end{eqnarray} where
\begin{equation}
u\equiv {\beta\over \theta_E}\; ; \; {u_1}={u}+{\pi \over
2}{\theta_E \over \theta_0}\; ; \;{u_2}={u}-{\pi \over 2}{\theta_E
\over \theta_0}. \label{b_dimlessTeVeS}
\end{equation} The magnification reduces to the
standard GR form as $\theta_0$ (or $r_0$) $\rightarrow \infty$. It
is then straightforward to study the light curves of
microlensing~\citep{pac86}.

            \begin{figure}[t]
\epsscale{1.7} \plottwo{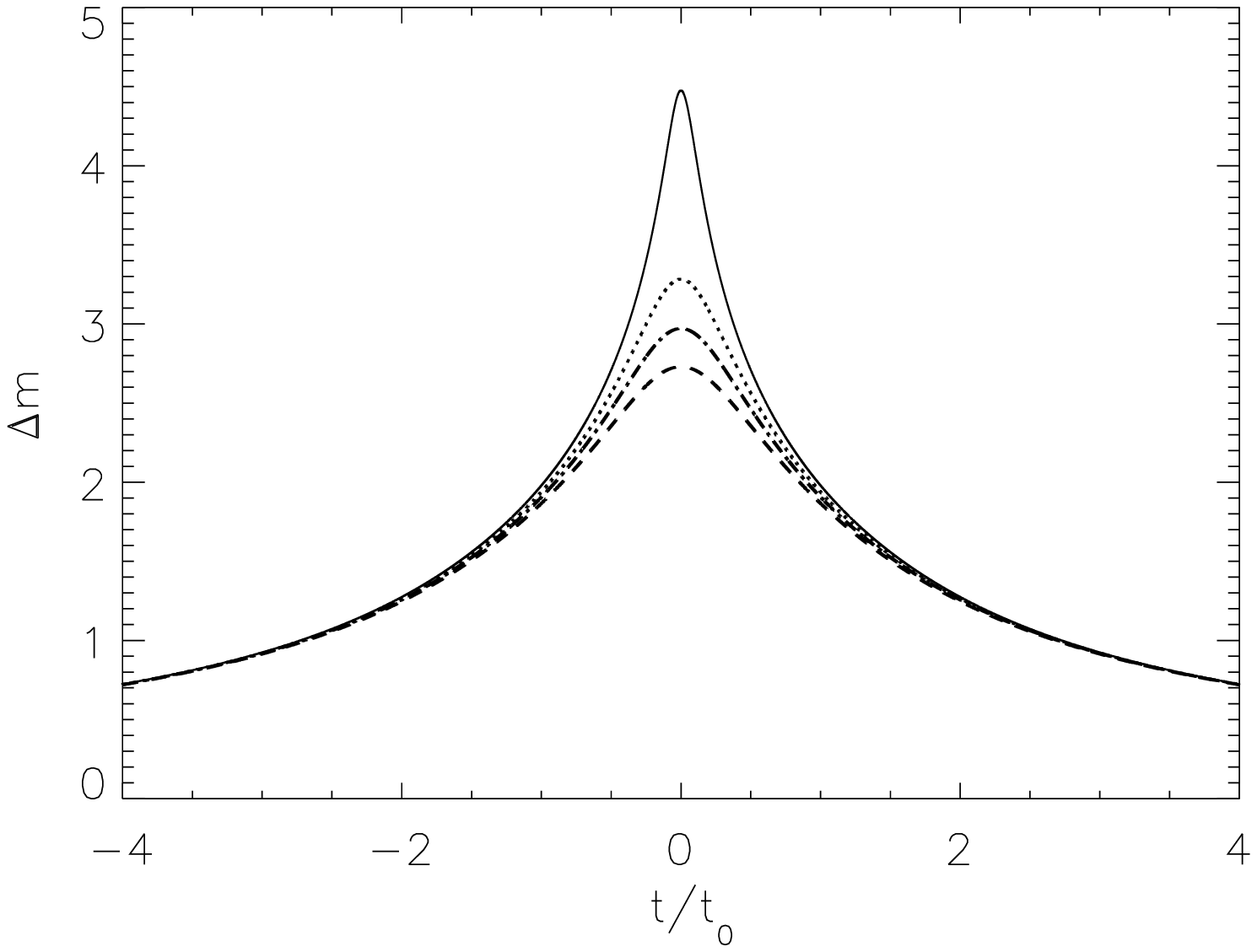}{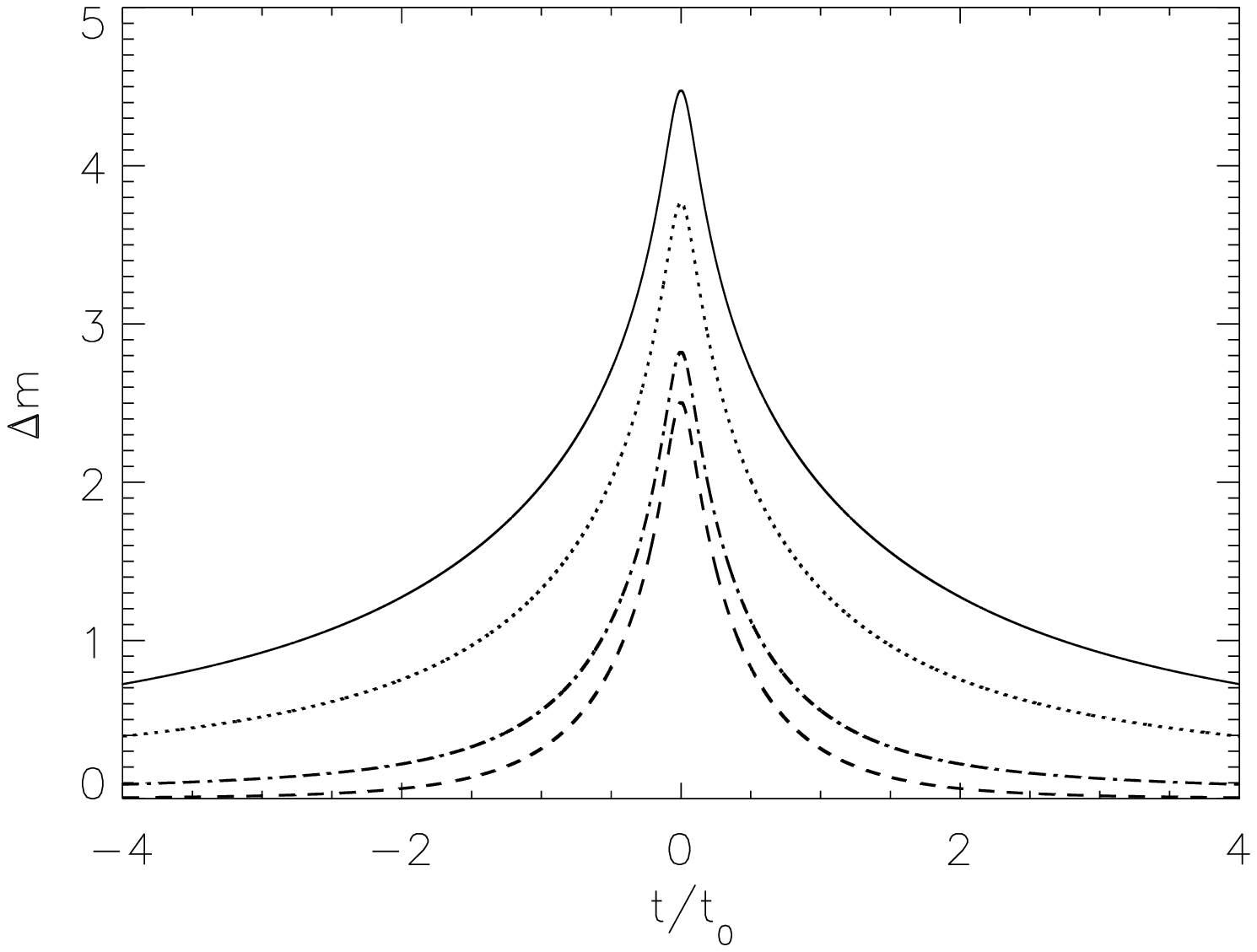}
\caption{\label{lightcurve} \textbf{upper panel:} Microlensing light
curves in T$e$V$e$S for $\theta_E /\theta_0 =$ 1.0. From solid line
to dashed line, $\beta_{min} /\theta_E =$ 0.1, 0.3, 0.4, 0.5.
\textbf{lower panel:} Light curves of microlensing in T$e$V$e$S with
$\theta_E /\theta_0 =$ 0.1 (dashed-dotted), 0.5 (dotted), 1.0
(solid) and in GR (dashed line). The difference of the magnitude is
given by $\Delta m = 2.5~\log A$ and $t_0$ is the time that sources
pass through Einstein radius.}
\end{figure}

Microlensing events results from the relative proper motions of
lenses to the sources. Because of proper motions, the alignment
between lens-observer and source-observer would vary with time,
\begin{equation}
u(t)=\sqrt{(\beta_{min}/\theta_E)^2+(t/t_0)^2}, \label{b_dimless}
\end{equation} and so does the magnification. Here $\beta_{min}$
occurs at the greatest of the alignment and $t_0$ is the time it
takes the source to move with respect to the lens by one $\theta_E$.
Putting Eq.~(\ref{b_dimless}) into Eq.~(\ref{magtotal}) and
Eq.~(\ref{b_dimlessTeVeS}), we can obtain the microlensing light
curves expressed in stellar magnitudes ($\Delta m \equiv 2.5\log
A$). Like the case in GR, the variation reaches the maximum as
$\beta(t) = \beta_{min}$, and for any given ${\pi}{{\theta_{_E}}^2
/2\theta_{_0}}$, the smaller the $\beta_{min}$, the greater
amplitude of the light curve. However, the shape of light curves in
T$e$V$e\,$S would deviate  from that in GR as ${\pi}{{\theta_{_E}}^2
/2\theta_{_0}}$ is significant (i.e., in deep MOND regime). In
general the contribution of ${\pi}{{\theta_{_E}}^2 /2\theta_{_0}}$
would raise the peak (see Fig.~\ref{lightcurve}).

Even though the differences in microlensing events between
T$e$V$e\,$S and GR only occurs at source redshift $z_s\geq
1$~\citep{mort1}, which is beyond the current micolensing projects
within the local group (see e.g.,~\citealt{alc00,pac96}),
observations such as several quasar microlensing events (see
e.g.,~\citealt{Wam01}) may help us to decide whether MOND is a
viable alternative to the dark matter paradigm.

    \section{Time Delay}\label{timedelay}
        Time-delay always plays an important role in gravitational lensing.
        It not only provides a more convenient way to get the lens
        equation~\citep{schn85,BN86} but also has some
        important values in its own right (for more complete reviews,
        see~\citealt{CSC02} or ~\citealt{Koch-Sch03}).
        As it is an alterative to the cold dark
        matter, it should be very interesting to study the behavior of time
        delay in T$e$V$e\,$S.

        \subsection{Arrival Time in T$e$V$e\,$S}\label{arrivaltime}
            We apply the static and isotropic metric described by
            Eq.~(\ref{weakspherical}) and assume that the photons are confined to the equatorial
            plane without loss of generality . Then, following the standard procedure
            of light bending (see e.g.~\citealt{wberg72}), not only the deflection angle can be derived, but also
            the equation which describes the arrival time of a light ray
            \begin{equation} \\(1-4\Phi)\dot{\varrho}^{2}=[{1-({1+4\Phi\over
            1+4\Phi_{0}})({\varrho_{0}\over\varrho})^{2}}]\dot{t}^{2}.
            \label{timeeq}
            \end{equation} This equation has a quadrature as a
            solution, which states that a photon from $\varrho_{0}$ to any distance
            $\varrho$ (or from $\varrho$ to the closest approach distance $\varrho_{0}$) would
            require a time,
            \begin{equation}
            \\t=\int^{\varrho}_{\varrho_{0}}{{(1-4\Phi)}^{1/2}d\varrho\over
            [1-(1+4\Phi)/(1+4\Phi_0)({\varrho_0\over\varrho})^{2}]^{1/2}}.
            \label{timeintegral}
            \end{equation} Taylor expanding Eq.~(\ref{timeintegral}) to the
            first order of $\mathcal{O}(\Phi)$, the original quadrature can be reduced to
            \begin{eqnarray}
            \lefteqn{t(\varrho,\varrho_{0})=\int^{\varrho}_{\varrho_{0}}{{d\varrho}
            {[1-({\varrho_0\over\varrho})^{2}]}^{-1/2}}} \nonumber \;\;\;\;\;\;\;\;\;\;\;\;\;\;\;\;\;\;\;\;\;\;\;\;\\
            \lefteqn{ +\int^{\varrho}_{\varrho_{0}}{d\varrho}{{2(\Phi-\Phi_0)(\varrho_0/\varrho)^{2}}\over
            {[1-({\varrho_0\over\varrho})^2]^{3/2}}}} \nonumber \;\;\;\\
            &&{-\int^{\varrho}_{\varrho_{0}}
            {d\varrho}{{2\Phi}\over{[1-({\varrho_0\over\varrho})^2]}{1/2}}}\;\;\;\;\;\;.
            \label{timeintegral2}
            \end{eqnarray}

            The second quadrature of Eq.~(\ref{timeintegral2}) differs from
            Eq.~(\ref{angle2}) only by a factor $\varrho_0$, whence, from this part, the
            arrival time contributions due to both gravitational mass and the scalar
            field are positive. However, this is not the case
            in the third quadrature of Eq.~(\ref{timeintegral2}). Since
            $\Phi_N$ and $\phi$ share an opposite sign in $\Phi$, their
            contributions to the resultant arrival time are also opposite. Moreover since
            the third quadrature is always larger than the second one
            (it is easily to observe this by multiplying by a factor ``$1- ({\varrho_0 /
            \varrho})^2$" in both numerator and denominator of the second
            quadrature),
            the time delay due to the scalar field will always be opposite to
            that due to the gravitational mass, i.e. with an opposite sign. In
            other words, the arrival time will be shorten rather than
            enhanced by the positive scalar field; this is a necessary
            condition in T$e$V$e\,$S for not violating causality~\citep{TeVeS}.
\begin{figure}[t]
\epsscale{1.0} \plotone{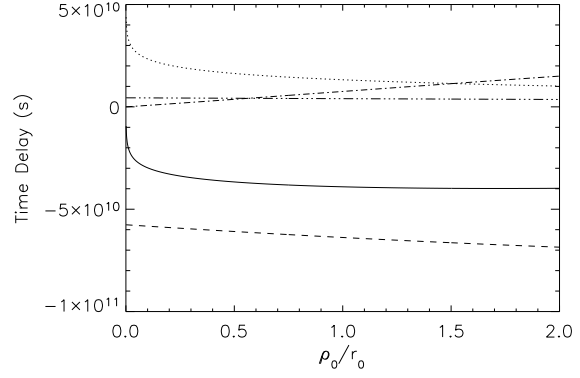} \caption{\label{timedelay1}
The time delay (the real light path minus the straight between
sources and observers) of a lensing system with a $M=5\times 10^{14}
M_\odot$ lens in the MOND regime. The closest approach is normalized
to the length scale $r_{_0}$ and both the observer and the source
are located at 10 $r_{_0}$. The dotted line and the dashed line show
the potential time delay induced by the ordinary matter and scalar
field respectively. Moreover the geometrical parts are represented
by the dotted-dashed (scalar field) and dotted-dotted-dashed line(
visible matter). Adding all components yields the total time delay
(solid line) and shows a negative effect on the arrival time in
T$e$V$e\,$S.}
\end{figure}

        \subsection{Arrival Time and the MOND Regime}\label{arrivaltimeMOND}
            As in the case of the deflection angle, in order to
            calculate time delay from Eq.~(\ref{timeintegral2}), we need some additional
            information on the free function, $F(\mu)$
            to determine the potential $\Phi$ in T$e$V$e\,$S.
            However, there are some differences in these two considerations.
            The first one arises from the fact that time delay is
            influenced by gravity even far away from the center. This makes
            a purely Newtonian regime unavailable. Therefore, we only consider
            the MOND regime here. Moreover, unlike the deflection angle case,
            we need information about $\Phi$, in addition to $\Phi'$ for the time delay.
            Eliminating $\mu$ of Eq.~(\ref{dscalar})
            from Eq.~(\ref{mu}) yields
            \begin{equation}
            \ \phi=(\Xi\mathfrak{a_0}Gm_g)^{1/2}\ln{4\pi\varrho \over kr_0} + \tilde{\phi}_{c},
            \label{scalar2}
            \end{equation} where $\tilde{\phi}_{c}$ can be absorbed into
            the symmetric physical metric $\tilde{g}_{\mu\nu}$ by rescaling
            the $t$ and $\varrho$ coordinates appropriately (Bekenstein
            2004).
            Recalling that $\Phi=\Xi\Phi_{N}+\phi$, and putting Eq.~(\ref{timeintegral2})
            into Eq.~(\ref{scalar2}), we get the arrival time in the MOND regime:
            \begin{eqnarray}
            \lefteqn{t(\varrho,\varrho_0)=(\varrho^2-\varrho_0^2)^{1/2}
            + 2\Xi Gm_g \left({\varrho-\varrho_0 \over \varrho+\varrho_{0}}
            \right)^{1/2}}\nonumber \\
            \lefteqn{{ +2\Xi Gm_g \ln({{\sqrt{\varrho^2-\varrho_0^2}}+\varrho \over
            \varrho_0})}}\nonumber \\
            \lefteqn{- 2 \left(\Xi\mathfrak{a}_0 Gm_g \right)^{1/2}} \nonumber \\
            & & \left[(\ln{({4\pi\varrho \over kr_0})}-1)(\varrho^2-\varrho_0^2)^{1/2}
            +\varrho_0 \sec^{-1}({\varrho \over \varrho_0})\right] \nonumber \\
            \lefteqn{+ 2 \left(\Xi\mathfrak{a}_0 Gm_g \right)^{1/2}} \nonumber \\
            & & \left[\varrho_0 \sec^{-1}({\varrho  \over \varrho_0})
            -\varrho_0^2{\ln({\varrho\over\varrho_0})}(\varrho^2-\varrho_0^2)^{-1/2}\right].
            \label{timeMOND}
            \end{eqnarray} Here again, $G$ differs by a factor of $\Xi+k/4\pi$ from the usual Newtonian
            constant, which is measured in
            the solar system experiments. We would like to stress that as a particular
            case of our cinsideration in $\S$ \ref{arrivaltime}, the sum of last two terms in
            Eq.~(\ref{timeMOND}) is always negative.

            Intrinsically, we may think that time delay (photon
            deviation of the path length) should increase with deflection
            angle as read off from the embedding diagram (Fig.\ref{figure0}).
            However, this is only true if we do not consider the
            deflection potential. Actually, a positive scalar field will offer
            a hyperbolic-like spacetime, in which the distance is
            shorter than that of a flat universe (see
            Fig.~\ref{timedelay1}).

        \subsection{Measurable Time Delay}

            Since it is the positions of sources and lens, not the distance
            from sources (or observers) to lens that are observed, time delay is
            conventionally expressed in terms of these dimensionless angles.
            It is appropriate to rewrite
            Eq.~(\ref{timeintegral2}) as:
            \begin{equation}
            \\t=\int^{\varrho}_{\varrho_{0}}dl - \int^{\varrho}_{\varrho_{0}} 2\Phi dl,
            \label{timeintegral3}
            \end{equation} here,
            $dl=[1-(1+4\Phi/1+4\Phi_0)({\varrho_0\over\varrho})^{2}]^{-1/2}d\varrho$ is
            a line element like that of Euclidean geometry, i.e.
            it obeys $dl^2=d\varrho^2+\varrho^2d\varphi^2$. Traditionally, we
            call the first quadrature in Eq.~(\ref{timeintegral3})
            the geometrical arrival time because it can be read off in a
            geometrical way (see Fig.~\ref{figure0}):
            \begin{eqnarray}
            t_{geom}(\varrho_{1},\varrho_{2})=\sqrt{(\varrho_0-\eta)^2+D_{LS}^2}
            +\sqrt{\varrho_0^2+D_{L}^2}.
            \label{timegeo}
            \end{eqnarray} Here $\eta=D_{S}\beta$. Indeed if we let a photon move from $\varrho_2$ to $\varrho_0$
            and from $\varrho_0$ to $\varrho_1$, it can be proved that the
            contributions to the arrival time calculated from the first two integrals
            of Eq.~(\ref{timeintegral2}) are identical to the geometrical arrival
            time, c.f. Eq.~(\ref{timegeo}). On the other hand, the third quadrature in
            Eq.~(\ref{timeintegral2}), which equals the second integral of
            Eq.~(\ref{timeintegral3}), contributes to what is called the \textit{potential
            time delay}. Adding
            the geometrical and potential contributions to the
            arrival time and subtracting the arrival time for an unlensed
            ray from $S$ to $O$, and considering the expansion of the cosmological background,
            we obtain the time delay of a possible ray compares to an un-deflected ray,
            \begin{equation}
            \triangle t=(1+z)\left[{D_{L}D_{S}\over 2D_{LS}}\big({\varrho_0\over D_L}-{\eta\over D_S}
            \big)^2-\psi(\varrho_0)\right]+ \mathbb{{B}}.
            \label{timedelayg}
            \end{equation} where, $\psi(\varrho_0)$ called the
            \textit{deflection potential} can be calculated from the potential time
            delay. Once again we only consider what happens in the MOND
            regime. For any given sources and lenses, the distances from
            the lenses to the observer (or sources) are decided, whence
            the potential time delay, which is given from the second and
            the fifth terms in Eq.~(\ref{timeMOND}), can be reduced to
            some constant plus the deflection potential
            contributions
            \begin{eqnarray}
            {\psi_{GR}= 4G_Nm_g \ln \varrho_{_0}}, \;\;\;\;\;\;\;\;\;\;\;\;\;\;\;\;\;\;\;
            \label{defpotGR} \\
            {\psi_{\phi}= 4 \left({\mathfrak{a}_0 G_Nm_g \over
            1+k/4\Xi\pi}\right)^{1/2}}
            \left({\pi \over 2}\varrho_0\right),
            \label{defpotscalar} \\
            {\psi_{corr}= -{4G_Nm_g \over 1+ 4\pi\Xi/k}
            \ln{\varrho_{_0}}}. \;\;\;\;\;\;\;\;\;\;
            \label{defpotcorr}
            \end{eqnarray}

\begin{figure}[t]
\plotone{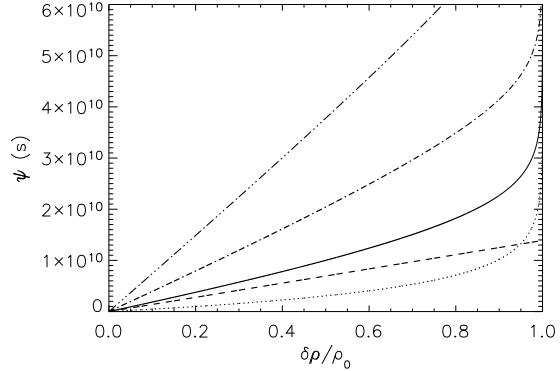}
\caption{\label{timedelay2} The measurable time delay (time delay
between two images for a given source) of a gravitational lensing
system in T$e$V$e\,$S. The system is assumed to have its largest
closest approach at $\varrho_{_0}=2\, r_{_0}$and a lens with $M=5\times
10^{14} M_\odot$ to resemble a cluster of galaxies alike Virgo
cluster. The dotted line is the contribution from the visible mass,
and the dashed line expresses the contribution from the scalar
field. Their combination is showed by the solid line. When
$\delta\varrho/\varrho_{_0}\rightarrow 0$, the discrepancy between
the time delay of the scalar field and ordinary matter decreases. On
the other hand, when $\delta\varrho/\varrho_{_0}\rightarrow 1$, the
influence of the ordinary matter increases and ultimately exceeds
that of the scalar field. For the same lens, two systems with
$\varrho_{_0}=5\,r_{_0}$ and $\varrho_{_0}=10\,r_{_0}$ respectively
are presented by dotted-dashed and dotted-dotted-dashed line.}
\end{figure}

            We have shown in \S~\ref{arrivaltimeMOND} that the time delay
            (relative to the undeflected path) is reduced rather enhanced in
            the T$e$V$e\,$S framework. Unfortunately, this phenomena cannot
            be measured in any GL system. For any given source at
            position $\beta$, we are only able to measure the time
            difference for two images $\varrho_{_0}^{(1)}$,
            $\varrho_{_0}^{(2)}$, which share the same lens potential,
            lens-observer, and lens-sources distances. In other words,
            the constant $\mathbb{B}$ in Eq.~(\ref{timedelayg}) is the same for all
            light rays from the source plane to the observer. Therefore
            in regard to the potential time delay, we can only measure
            the contributions from the deflection potential, which possesses the
            same sign in GR and T$e$V$e\,$S (Fig.~\ref{timedelay2}). Moreover, since the deflection potential
            does not depend on the length scale $kr_{_0}/4\pi$, the free choice of $\hat{\phi}_c$
            has no influence on the measurable time delay in GL systems.
            \begin{figure}[t]
\plotone{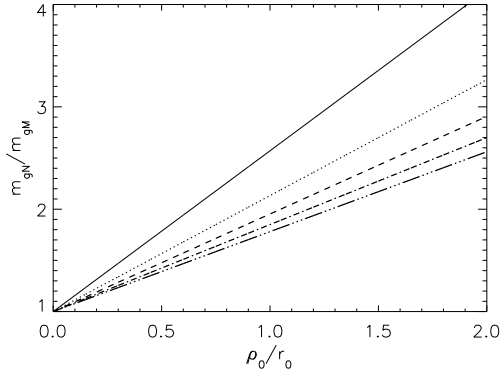} \caption{\label{figure5} Mass ratios of the
MOND and dark matter paradigms. The solid line shows the constraint
on the mass ratio while requiring an identical amount of deflection
angle in the two different paradigm. This constrain is found again
from the requirement of the measurable time delay as
$\delta\varrho\ll\varrho_{_0}$. However, the corresponding mass
ratios of the time delay will be increasingly less than that of the
deflection angle as the difference between the two closest approach
increases. From the dotted line to the dotted-dotted-dashed line,
$\varrho_{_0}^{1}/\varrho_{_0}^{2}=$ 2 ,3, 4, 5.}
\end{figure}

            Even though we cannot measure the opposite contribution to the
            arrival time in T$e$V$e\,$S, time delay between two images due to
            the deflection potential offers another constraint on the mass ratio, which is
            not a $priori$ identical to that due to deflection angle.

            In order to illustrate this, we consider a general case:
            the value of $\varrho_{_0}^{_1}$ is ``${n}$" times
            $\varrho_{_0}^{_2}$, whence the time delay between the two
            images is
            \begin{equation}
            \Delta\psi= 4G_Nm_g \left[ \ln n+
            {\pi \over 2}{(n-1) \over n} {\varrho_{_0}^{_1} \over r_{_0}} \right].
            \label{timedelayimag}
            \end{equation}

            For the extreme case, the difference between $\varrho_{_0}^{_1}$ and
            $\varrho_{_0}^{_2}$ is negligible compared to either
            distance. Hence, $ \ln (\varrho_{_0}^{_1}/\varrho_{_0}^{_2}) \simeq
            \delta\varrho_{_0} /\varrho_{_0}$, and
            Eq.~(\ref{timedelayimag}) can be reduced to:
            \begin{equation}
            \Delta\psi= 4G_Nm_g {\delta\varrho_{_0} \over \varrho_{_0}}
            \left( 1+
            {\pi \over 2} {\varrho_{_0} \over r_{_0}} \right).
            \label{timedelayimag2}
            \end{equation}
            If we compare Eq.~(\ref{timedelayimag2}) with
            Eq.~(\ref{alphatheta}), we can find  that the similarity of the two equations
            makes the mass ratio $m_{gN}/m_{gM}$ exactly the same under the requirement of
            either identical deflection angle or time delay. However, if
            the difference between $\varrho_{_0}^{_1}$ and
            $varrho_{_0}^{_2}$ is not small, the mass ratio given by the time delay will
            be less than that obtained from the deflection angle
            (Fig.~\ref{figure5}).

        \subsection{Time Delay and Lensing Equation}
            As showed by Schnieder (1985) and Blandford and Narayan (1986),
            time delay with Fermat's principle offers an alternative for
            deriving
            the lens mapping. Moreover, since Fermat's
            principle has proved valid
            under a very general metric~\citep{Kovner90,per90}, we should be able to re-derive lens equation from our time
            delay result.
            Fermat's principle asserts that a light path is true if and
            only if its arrival time or, what amounts to the same, its time
            delay is stationary with respect to variation of the turning
            point. In other words, $\partial(\triangle t)/\partial
            \varrho_{_0}=0$. With this sufficient and necessary condition
            for
            Fermat's principle, we can obtain the lens equation in the MOND
            regime,
            \begin{equation}
            \ {\eta}={D_S \over D_L}{\varrho_{_0}}-D_{LS}{\nabla \psi}.
            \label{lenseq2}
            \end{equation} $\nabla \psi$ is composed of
            three parts; and when $\varrho$ is very large
            \begin{eqnarray}
            \nabla \psi_{GR}= 4 {{G_Nm_g } \over \varrho_{_0}}
            \;\;\;\;\;\;\;\;\;\;\;\;\;\;\;\;\;\,
            \\
            \nabla \psi_{\phi}= 2{\pi} \left({\mathfrak{a}_0 G_Nm_g \over 1+k/4\Xi\pi}\right)^{1/2}\;
            \\
            \nabla  \psi_{corr}= -4 \left( {G_Nm_g \over
            {1+4\pi\Xi/k}}\right) .
            \end{eqnarray}
            Obviously the addition of the last three equations equals
            $\Delta\varphi$, given by Eq.~(\ref{lenform}), hence Eq.~(\ref{lenseq2}) is identical to the lens equations
            given by Eq.~(\ref{lenequ}).

    \section{Discussion}\label{discussion}
        Since the appearance of Bekenstein's relativistic MOND
        gravitation, the long-time incompleteness of Milgrom's modified
        Newtonian dynamics seems to be filled up. Now we can investigate
        the GL phenomenon, such as the deflection angle or time delay, in
        the MOND paradigm, which could not be done before.

        In this paper,we have investigated the GL phenomenon under the
        approximations and presuppositions of

        ($i$) The GL lens is assumed to be static, spherically symmetric and
        following the thin lens formalism,

        ($ii$) The motions of light rays are described in the framework of a
        Schwarzschild lens (i.e. a point mass model),

        ($iii$) The revised physical metric is obtained by adding a positive
        scalar field into the potential of the standard Schwarzschild metric in symmetric
        coordinates,


        Under these presuppositions, we find that when $\theta$ is larger than $\theta_{_0}$
        ($\equiv r_{_0}/D_L$; the MOND length scale),
        the reduced deflection
        angle $\alpha$ will approach to a constant for a given mass. This special prediction
        is a feature of GL in T$e$V$e\,$S, which is similar to that obtained
        by~\citet{mort1} with an intuitional approach. This is no surprise
        because just like in GR, the deflection of photons is simply twice
        the deflection of a massive particle with the speed of light in T$e$V$e\,$S, and
        thus~\citet{mort1} started from the correct premise. However, there
        is still something that was unknown before the appearance of T$e$V$e\,$S. For
        a static spherically symmetric spacetime, the case $\mu\ll1$,
        which yields a relation of Eq.~(\ref{mu}), is valid only when
         $|\mbox{\boldmath{$\nabla$}}\Phi|\ll (4\pi/k)^2\mathfrak{a}_0$. This condition
         is valid in the MOND regime when $|\mbox{\boldmath{$\nabla$}}\Phi|$
        goes up to a couple of orders above $\mathfrak{a}_0$, or
        equivalently $\varrho_0$ is around one order of magnitude
        below $r_0$.~\citep{TeVeS}

        We should address that the criteria of mass discrepancy
        for GL effects in T$e$V$e\,$S consists with that of stellar dynamics.
        In other words, only when $\varrho_{_0}\gg kr_{_0}/4\pi$, the
        ``missing mass" shall appear. This corresponds to the demarcation of
        the high surface brightness (HSB) and low surface brightness {LSB} galaxies from
         the dynamical analysis~\citep{SandersMcGaugh}.

         When we apply the deflection angle law to magnification, we find that in
         T$e$V$e\,$S the difference in the magnifications
         of the two images in the
         point mass model depends on the lens mass and source positions,
         and is always larger than one. This differs
         from traditional gravitational
         lensing, which says that the difference must always
         be one. Tens of thousands of the multiple images lensings found by the Sloan
         Digital Sky Survey (SDSS)\citep{stoueal} might be applied to check
         this prediction. For microlensing,
         light curves in T$e$V$e\,$S at
         deep MOND regime also differ from that in GR. To observe the discrepancy, the
         sources have to be located about $z_{s}\geq 1$~\citep{mort1}.

        Concerning time delay, the result is even more exotic.
        Add an arbitrary
        positive scalar field into the primary Schwarzschild
        metric~\citet{TeVeS}
        the resultant contributions of the scalar field will reduce
        rather than enhance the potential time delay. Unfortunately,
        this phenomenon is unmeasurable in GL systems. What we can
        determine is only the time delay between two images produced
        by the same source. Therefore, the opposite feature of the time
        delay in T$e$V$e\,$S, which is the same for all images, will be
        canceled out, and only those parts contributed from the
        deflection potential can be observed.

        Even though the opposite effect on the potential time delay due to the scalar
        field can not be measured in GL system, the time delay between
        two images offers another constraint on the needed mass in GR and
        T$e$V$e\,$S, which usually differs from that given by the deflection angle.
        Actually the mass ratio obtained from
        these two approaches will be the same, otherwise the mass ratio
        given by time delay would always be smaller than that by deflection angle.
        In one word, the MOND
        and CDM paradigm are not mutually alternatives in a GL system,
        when we consider deflection angle and time delay at the same
        time.

        The first time delay for the gravitational lens Q0957+561
        was measured in 1984~\citep{FN84}, since then more than 11 time delay lenses have
        been found, including 7 systems with a good quality of the
        astrometric data and 2 systems with serious problems~\citep{Koch-Sch03}. It would be
        very interesting if those debatable systems can be explained in
        the framework of T$e$V$e\,$S. However, we  emphasize that
        the measurement of time delay is controversial.
        Actually it took almost 20 years of debate between a short
        delay and a long delay on the first observed time delay
        source~\citep{PLW01}.

        It also worth noting that our conclusion is based on the
        Bekenstein's approach to investigate a gravitational lens. He
        assumes a $priori$ that the potential dominating in the massive
        particle dynamics agrees with the potential influences on the GL
        system. In other words, the temporal and spatial components of
        the metric respectively have a form $-(1+2\Phi)$ and $(1-2\Phi)$,
        whence $\tilde{g}_{tt}\tilde{g}_{ii}\simeq 1$. However, there is
        still a chance that these two potentials are not the same and
        even that
        $\tilde{g}_{tt}\tilde{g}_{ii} \neq 1$ under the framework of
        T$e$V$e\,$S. If so, the conclusion of this paper may be
        different~\citep{Edery,BMS00}.

        Since we can observe the influence of gravity on gravitational
        lens phenomena at distances further away than we can for the dynamics of massive
        particles, GL systems offer a good method to distinguish between GR and
        T$e$V$e\,$S. Although we need more than the point mass model to
        fit astrometric data, this simplified model has in principal
        told us the exotic predictions of the deflection angle and time
        delay. It would be very interesting if we could develop a more
        realistic model to fit the known data in GL systems.







    \acknowledgements   
   We thank J.M. Nester for stimulating discussions on the modified gravitational theories
   and comments on the whole work. We
   also thank Chi Yuan, Tzi-Hong Chiueh,  Kin-Wang Ng and Pisin Chen for the comments.
   We appreciate the referee for pointing out the potential
   observation of microlensing.
   M.C. Chiu and C.M. Ko are supported in part by the National Science
   Council of Taiwan, by grant NSC-93-2112-M-008-017. Y. Tian is
   partially supported by the grant NSC93-2112-M-002-025 of  the National Science Council.
   Y. Tian wishes to thank Prof. W-Y. P. Hwang for
   the supervision during which this research was carried out.


\begin{thebibliography}{}
      \bibitem[Alcock et al.(2000)]{alc00}{Alcock C., et al. 2000, \apj, \textbf{542}, 281}
       \bibitem[Bekenstein(1988)]{phase}{Bekenstein, J.D.} 1988, Phys. Lett. B,
           \textbf{202}, 497
       \bibitem[Bekenstein(2004)]{TeVeS}{Bekenstein, J.D.} 2004,  \prd,
           \textbf{70}, 083509

       \bibitem[Bekenstein \& Milgrom(1984)]{BekMilg}{Bekenstein, J.D. \& Milgrom, M.} 1984,
         \apj, \textbf{286}, 7
       \bibitem[Bekenstein et al.(2000)]{BMS00}Bekenstein, J.D.,
            Milgrom, M. \& Sanders, R.H. 2000, \prl, \textbf{85}, 1346
       \bibitem[Bekenstein \& Sanders(1994)]{BekSan}Bekenstein, J.D.  \& Sanders, R.H. 1994,
           \apj, \textbf{429}, 480

       \bibitem[Begman(1989)]{Bgmn89} {Begman, K.G. } 1989,
          A$\&$A, \textbf{223}, {47}
        \bibitem[Begeman et al.(1991)]{BBS} {Begeman, K.G., Broeils, A.H. \&
          Sanders, R.H.} 1991, \mnras, \textbf{249}, 523
        \bibitem[Blandford \& Narayan(1986)]{BN86}{Blandford, R.D. \& Narayan,
          R.} 1986, \apj, \textbf{310}, 568
        \bibitem[Blandford \& Narayan(1992)]{BN92}{Blandford, R.D. \& Narayan,
          R.} 1992, \araa, \textbf{30}, 311
        \bibitem[B\"{o}hringer(1995)]{Bohringer95} {B\"{o}hringer, H.}
          1995, RevMexAA, \textbf{8}, {259}
        \bibitem[Courbin et al.(2002)]{CSC02}{Courbin, F., Saha,
          P. \& Schechter, P.L.} 2002, Gravitational Lensing: An Astrophysical Tool, Edited by F. Courbin
               D. Minniti, Lecture Notes in Physics, Vol. 608 (astro-ph/0208043)
        \bibitem[Dyer \& Roder(1973)]{dyer73}{Dyer, C.C. \& Roder,
          R.C.} 1973, \apj, \textbf{180}, {31}
        \bibitem[Edery(1999)]{Edery}{Edery, A.} 1999, \prl,  \textbf{83}, 3990
        \bibitem[Fabbiano et al.(1989)]{Fabb89}
           {Fabbiano, G.F., Gioia, I.M. \& Trinchieri, G.} 1989, \apj,
           \textbf{347}, {127}
        \bibitem[Florentin-Nielsen(1984)]{FN84}{Florentin-Nielsen,
          R.} 1984, A$\&$A, \textbf{138}, {L19}
        \bibitem[Giannios (2005)]{gian05}{Giannios, D} 2005, \prd, \textbf{71}, {103511}
        \bibitem[Hao \& Akhoury(2005)]{hao05}{Hao, J.G. \& Akhoury, R.} 2005, preprint (astro-ph/0504130)
        \bibitem[Kauffmann et al.(1993)]{kauff93}
          {Kauffmann, G., White, S. D. M., \& Guiderdoni B.} 1993, \mnras, \textbf{264}, {201}
        \bibitem[Kazantzidis et al.(2004)]{kaza04} {Kazantzidis, S., Mayer, L., Mastropietro, C., Diemand, J.,
          Stadel, J., \& Moore, B.} 2003, \apj, \textbf{611}, {L73}
        \bibitem[Kochanek(2004)]{Koch04} {Kochanek, C.S.} 2004,
          astr-ph/0412089
        \bibitem[Kochanek \& Schechter(2003)]{Koch-Sch03}{Kochanek, C.S. \&
        Schechter, P.L.} 2003, preprint (astr-ph/0306040)
        \bibitem[Kovner(1990)]{Kovner90}{Kovner, I.} 1990, \apj, \textbf{351}, {114}

        \bibitem[Loewenstein \& White(1999)]{Loew99} {Loewenstein, M. \&
          White, R.E.} 1989, \apj, \textbf{518}, {50}
        \bibitem[Maller \& Dekel(2002)]{Maller02} {Maller, A.H. \& Dekel, A.}
          2002, \mnras, \textbf{335}, {487}
        \bibitem[McGaugh(1999)]{mc99}{McGaugh, S.S.} 1999, \apj, \textbf{523}, {L99}
        \bibitem[McGaugh(2004)]{mc04}{McGaugh, S.S.} 2004,
          astr-ph/0312570
        \bibitem[McGaugh \& de Blok(1998)]{mcdb98} {McGaugh, S.S. \& de Blok, W.J.G.} 1998,
          \apj, \textbf{499}, 66
        \bibitem[Milgrom(1983a)]{M1} {Milgrom, M.} 1983a, \apj, \textbf{270}, {365}

         \bibitem[Milgrom(1983b)]{M2} {Milgrom, M.} 1983b \apj, \textbf{270}, {371}

        \bibitem[Milgrom(1983c)]{M3} {Milgrom, M.} 1983c, \apj,  \textbf{270}, {384}
        \bibitem[Milgrom \& Sanders(2003)]{MS03}{Milgrom, M. \& Sanders, R.H.} 2003,
          \apj,  \textbf{599}, {L25}
        \bibitem[Moore et al.(1999)]{moore99} {Moore, B., Ghigna,
          S., Governato, F., Lake, G., Quinn, T., Stadel, J., \&
          Tozzi, P. } 1999, \apj, \textbf{524}, {L19}
        \bibitem[Mortlock \& Turner(2001)]{mort1} Mortlock, D.J. \& Turner, E.L. 2001, \mnras
          \textbf{327}, 557
        \bibitem[Navarro et al.(1995)]{NFW95} {Navarro J.F., Frenk C.S. \& White S.D.M.}
          1995, \apj, \textbf{275}, {56}
        \bibitem[Navarro \& Steinmetz(2000)]{NS00} {Navarro, J.F. \& Steinmetz, M.}
          2000, \apj, \textbf{528}, {607}
        \bibitem[{Paczy\'{n}ski }(1986)]{pac86}{Paczy\'{n}ski B. 1986, \apj, \textbf{304}, 1}
        \bibitem[Paczy\'{n}ski(1996)]{pac96}{Paczy\'{n}ski B. 1996, \araa, \textbf{34}, 419}
        \bibitem[Perlick(1990)]{per90}{Perlick, V.} 1990, Class.
          Quant. Grav., \textbf{7}, 1319
        \bibitem[Petters, Levine \& Wambsganss(2001)]{PLW01}
          {Petters, A.O., Levine, H. \& Wambsganss, J.} 2001, Singularity Theory
          and Gravitational Lensing (Boston: Birkh\"{a}user)
        \bibitem[Romanowsky et al.(2003)]{Rosky03} Romanowsky, A.J., Douglas, N.G., Arnaboldi, M., Kuijken, K., Merrifield, M.R.,
Napolitano, N.R.,  Capaccioli, M., \&  Freeman, K.C. 2003, Science, \textbf{301}, 1696

        \bibitem[Sanders(1988)]{SPCG}{Sanders, R.H.} 1988, \mnras
          \textbf{235}, 105
        \bibitem[Sanders(1996)]{rhs96} {Sanders, R.H.} 1996, \apj,
          \textbf{473}, {117}
        \bibitem[Sanders(1997)]{rhs97} Sanders, R.H. 1997, \apj, \textbf{480}, 492
        \bibitem[Sanders \& McGaugh(2002)]{SandersMcGaugh} {Sanders, R.H. \& McGaugh, S.S.}
        2002, \araa, \textbf{40},  263
        \bibitem[Sanders \& Verheijen(1998)]{sv98} {Sanders, R.H. \& Verheijen, M.A.W.} 1998,
         \apj, \textbf{503}, 97
        \bibitem[Schneider(1985)]{schn85}{Schneider, P.} 1985, A$\&$A, \textbf{143}, 413
        \bibitem[Schneider(1996)]{schn96}{Schneider, P.} 1996, IAUS, \textbf{168}, {209}
        \bibitem[Schneider et al.(1992)]
          {schn-ehl-fal92}{Schneider, P., Ehlers, J., \& Falco,
          E.E.} 1992, \textit{Gravitational Lenses} (New York: Springer-Verlag)
        \bibitem[Skordis et al.(2005)]{SMFB05} {Skordis, C., Mota, D.F.,
          Ferreira, P.G., \& Boehm, C.} 2005, preprint (astro-ph/0505519)
        \bibitem[Smith(1936)]{Smith} {Smith, S.} 1936, \apj, \textbf{83}, {23}
        \bibitem[Soussa \& Woodard(2003)]{Soussa1}Soussa, M.E., \& Woodard, R.P. 2003, Class. Quant.
        Grav.,
          \textbf{20}, 2737

        \bibitem[Soussa \& Woodard(2004)]{Soussa2}Soussa, M.E., \& Woodard, R.P. 2003, Phys. Lett.
        B,
          \textbf{578},  253
       \bibitem[Spergel et al.(2003)]{spe03}{Spergel, D.N. et al.}
         2003, \apjs ,\textbf{148}, {175}
       \bibitem[Stoughton et al.(2002)]{stoueal}{Stoughton, C et al.}
         2002, \aj ,\textbf{123}, {485}
       \bibitem[van Albada et al.(1985)]{vanAlba85} {van Albada, T.S.,
       Bahcall, J.N., Begeman, K. \& Sanscisi, R.} 1985, \apj, \textbf{295}, {305}
       \bibitem[Wambsganss(2001)]{Wam01}
          {Wambsganss, J.} 2001, PASA, \textbf{18}, {207}
       \bibitem[Weinberg(1972)]{wberg72}{Weinberg, S.} 1972,
          \textit{Gravitation \& Cosmology} (New York: Wiley)
       \bibitem[Weyl(1923)]{weyl23}{Weyl, H.} 1923, Z. Phys.,
         \textbf{24}, 230
       \bibitem[Zhao (2005a)]{zhao05a}{Zhao H.S.} 2005a, preprint(astro-ph/0508635)
       \bibitem[Zhao (2005b)]{zhao05b}{Zhao H.S.} 2005b, A$\&$A Lett. accepted
       \bibitem[Zhao et al.(2005)]{zhaoeal05}{Zhao H.S. et al.} 2005,
         \mnras~ submitted (astro-ph/0509590)
        \bibitem[Zwicky(1933)]{Zwicky} {Zwicky, F.} 1933, \textit{Helv. Phys. Acta} \textbf{6}, {110}



    \end{thebibliography}
\end{document}